\documentclass[reprint, superscriptaddress, secnumarabic,amssymb, nobibnotes, aps, prl]{revtex4-2}

\usepackage{graphicx}
\usepackage{epstopdf}
\usepackage[T1]{fontenc}
\usepackage[latin9]{inputenc}
\usepackage{amsbsy}
\usepackage{gensymb}
\usepackage[T1]{fontenc}
\usepackage[latin9]{inputenc}
\usepackage{amsmath}
\usepackage{amssymb}
\usepackage{bbm}
\usepackage{braket}
\usepackage{nameref}
\usepackage{xcolor}
\allowdisplaybreaks
\usepackage{graphicx}
\usepackage[colorlinks=true]{hyperref}
\usepackage{orcidlink}
\hypersetup{
    bookmarks=true,         
    unicode=false,          
    pdftoolbar=true,        
    pdfmenubar=true,        
    pdffitwindow=false,     
    pdfstartview={FitH},    
    pdftitle={},    
    pdfauthor={R. K. Kushwaha, R. P. Singh},     
    pdfsubject={},   
    pdfcreator={},   
    pdfproducer={}, 
    pdfkeywords={} {} {}, 
    pdfnewwindow=true,      
    colorlinks=true,       
    linkcolor=blue, 
    citecolor=blue,        
    filecolor=blue,      
    urlcolor=blue           
}
\usepackage[normalem]{ulem}

\renewcommand{\approx}{\simeq}

\newcommand{\RNum}[1]{\uppercase\expandafter{\romannumeral #1\relax}}

\usepackage{graphicx}
\usepackage{epstopdf}
\usepackage[T1]{fontenc}
\usepackage[latin9]{inputenc}
\usepackage{amsbsy}
\usepackage{gensymb}

\begin{document}

\title{\textrm{High Critical Temperature and Field Superconductivity in Nb$_{0.85}$X$_{0.15}$, (X = Ti, Zr, Hf) Alloys: Promising Candidates for Superconducting Devices}}
\author{{R.~K.~Kushwaha}\,\orcidlink{0009-0005-3457-3653}}
\affiliation{Department of Physics, Indian Institute of Science Education and Research Bhopal, Bhopal, 462066, India}
\author{{S. Jangid}}
\affiliation{Department of Physics, Indian Institute of Science Education and Research Bhopal, Bhopal, 462066, India}
\author{{P. Mishra}}
\affiliation{Department of Physics, Indian Institute of Science Education and Research Bhopal, Bhopal, 462066, India}
\author{{S. Sharma}}
\affiliation{Department of Physics, Indian Institute of Science Education and Research Bhopal, Bhopal, 462066, India}
\author{{R.~P.~Singh}\,\orcidlink{0000-0003-2548-231X}}
\email[]{rpsingh@iiserb.ac.in}
\affiliation{Department of Physics, Indian Institute of Science Education and Research Bhopal, Bhopal, 462066, India}
\date{\today}

\begin{abstract}
\begin{flushleft}
\end{flushleft}
Niobium and its alloys with early transition metals have been extensively studied for their excellent superconducting properties. They have high transition temperatures, strong upper critical fields, and high critical current densities, making them ideal for superconducting applications such as SQUIDs, MRI, NMR, particle accelerators, and Qubits. Here we report a systematic investigation of as-cast Nb-rich alloys, Nb$_{0.85}$X$_{0.15}$ (X = Ti, Zr, Hf), using magnetization, electrical transport, and specific heat measurements. They exhibit strong type-II bulk superconductivity with moderate superconducting transition temperatures and upper critical fields. The estimated magnetic field-dependent critical current density lies in the range of 10$^5$--10$^6$~A/cm$^2$ across various temperatures, while the corresponding flux-pinning force density is on the order of GNm$^{-3}$, suggesting the potential of these materials for practical applications. Electronic-specific heat data reveal a strongly coupled, single, isotropic, nodeless superconducting gap. These Nb-rich alloys, characterized by robust superconducting properties, hold significant potential for applications in superconducting device technologies.
\end{abstract}

\maketitle

\section{Introduction}
Superconductivity is one of the most fascinating phenomena in condensed matter physics, offering both deep insights into quantum states of matter and broad technological potential. Among various superconducting materials, niobium (Nb)-based superconductors have emerged as highly sought-after materials in research due to their potential applications, characterized by high superconducting transition temperatures ($T_c$) and upper critical fields ($H_{c2}$)~\cite{coombs2024high,hampshire1974critical,largascaleapplication,priinits2024peculiarities}. Extensive studies have been conducted on Nb-based A-15~\cite{STEWART201528,Nb3SnPRM,Kinoshita01081990,rodrigues2000development,Nb3Al_recent} and Nb-T (T: early transition metals)~\cite{NbX} superconductors~\cite{PhysRevMaterials.8.084801}. These materials show a wide array of physical properties, including large magnetoresistance~\cite{LargeMRNb3Sb}, non-trivial band topology~\cite{NTSS}, and their ability to form wires~\cite{Nb3SnPRM,rodrigues2000development} and superconducting joints~\cite{banno2021high,patel2019niobium}. Such characteristics make them promising candidates for advanced technologies~\cite{seeber1998commercially,seeber1998power,kalsi2003installation}. In addition to industrial applications, Nb and Ta are widely recognized as suitable materials for superconducting radio frequency (SRF) cavities~\cite{Nbcavity} and Josephson junctions~\cite{NbRF1,NbRFsurface}, which facilitate the development of superconducting qubits~\cite{qbits1,qbits2}. Recent studies have shown that Ta-Zr~\cite{Ta-Zr} and Ta-Hf~\cite{Ta-Hf} alloys also exhibit promising superconducting properties. Owing to the high electropositivity of Hf and Zr, their alloys are considered strong candidates for meeting theoretical benchmarks aimed at enhancing qubit performance. Nb and Ta-based alloys, characterized by diverse crystal structures and physical properties, can provide a foundation for exploring similar compounds.

Among Nb-T alloys, Nb-Ti alloys~\cite{NbTihon2003composition,NbTiPhysRevB.110.L140502,baker1969correlation,jablonski1993niobium} have achieved widespread commercial success in superconducting magnets and have also recently been studied under high pressure, setting new records for both $T_c$ (= 19.1~K) and $H_{c2}$ (= 19~T) among all alloys composed only of transition metals~\cite{NbTipressure}. 
\begin{figure*} [t] 
\includegraphics[width=1.35\columnwidth,origin=b]{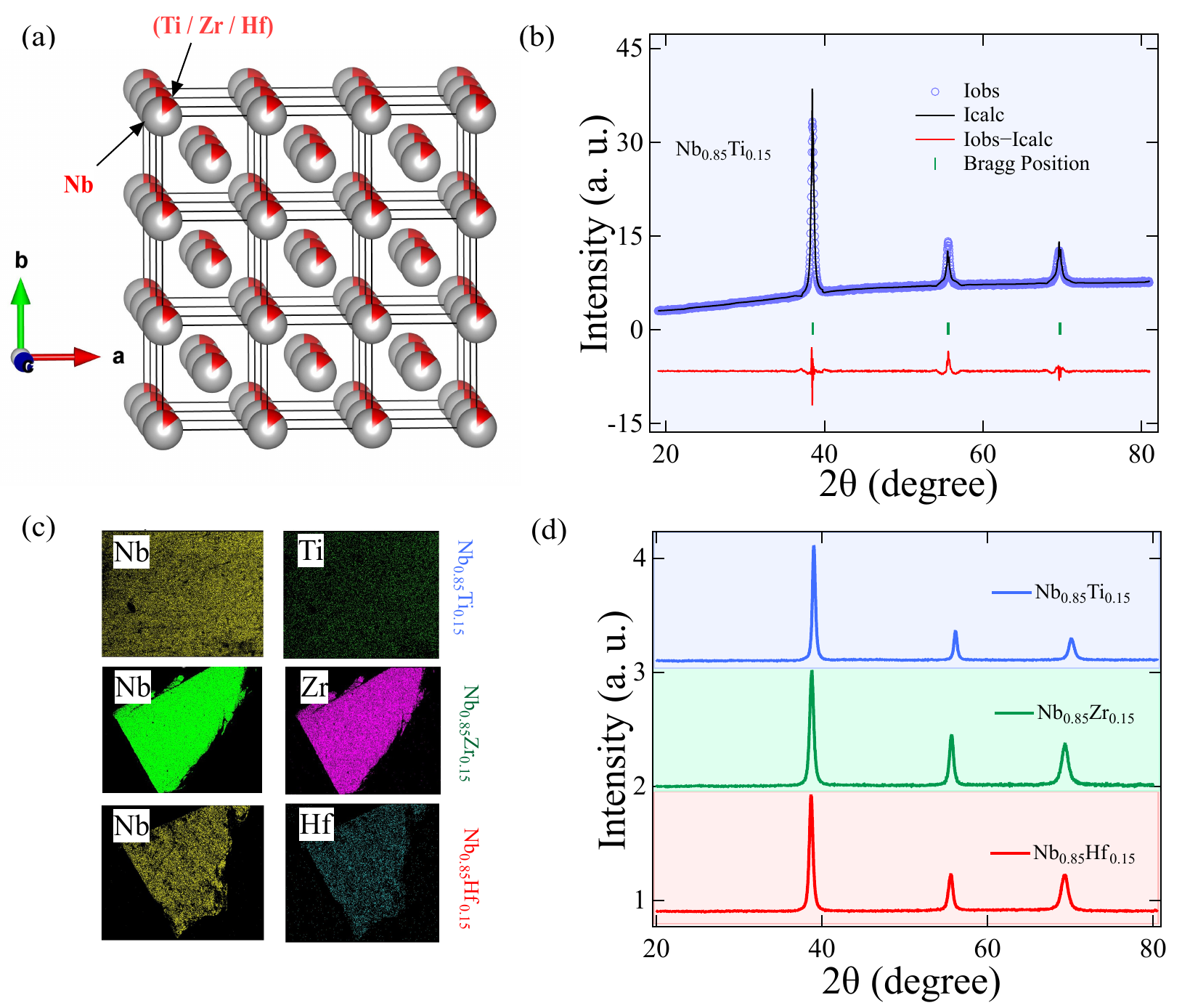}
\caption{\label{XRD} (a) The bcc crystal structure of Nb$_{0.85}$X$_{0.15}$, (X = Ti, Zr, Hf) (b) Rietveld refinement of powder XRD pattern of Nb$_6$Ti alloy (c) Elemental mapping of  Nb$_{0.85}$X$_{0.15}$ (X = Ti, Zr, Hf) alloys (d) Room temperature XRD patterns of Nb$_{0.85}$X$_{0.15}$, (X = Ti, Zr, Hf).}
\end{figure*}
Additionally, Nb-Ti alloys with Nb content 50$\%$ or more are applicable in osseous implant devices, such as orthopedic and orthodontic implants~\cite{TixNb}. Furthermore, Nb-Zr~\cite{NbZrPRM,Nb0.75Zr0.25,fietz1964magnetization,NbZrPhysRevLett.6.671,mirmefstein1997mixed,junod2002specificNb0.77Zr0.23} and Nb-Hf~\cite{Nb-Hf,Nb-Hfconf,Nb-Hfkoch1979peak} alloys are also decisive in the same context. Nb-Zr alloys also present significant advantages for SRF cavities, including higher transition temperature, which facilitates reduced energy dissipation and cryogenic costs. Their simplified phase diagrams enable easier processing and optimization, while a stable ZrO$_2$ native oxide enhances RF performance by minimizing quench-inducing defects~\cite{ZrdopedNb}. In addition, the combination of high values $T_c$ and $H_{c2}$ with low surface resistance, the absence of nodes in the gap symmetry, metallic behavior, high thermodynamic critical field and superheating field, and material morphology ~\cite{valente2016superconducting} make these alloys excellent candidates for high-performance SRF cavities. However, most of the research on these Nb-based alloys has focused on the A15 phase; a detailed study of the cubic $\alpha$-W structure, which is the same as pure niobium, is lacking. These alloys are expected to exhibit superior superconducting and physical properties, making them suitable for superconducting device applications. Therefore, a detailed study of niobium-rich alloys in $\alpha$-W structure is crucial to identify promising candidates for practical applications.

In this paper, we have investigated the superconducting and normal-state properties of the Nb-rich family of alloys, Nb$_{0.85}$X$_{0.15}$ (nominal composition: Nb$_6$X, X = Ti, Zr, Hf). We substitute elements Ti (3d), Zr (4d), and Hf (5d) at the X site with a fixed Nb content. We perform a comparative analysis of superconducting and normal-state parameters derived from electrical resistivity, magnetization, and specific heat measurements, which remain relatively unexplored in the literature. Temperature-dependent magnetization loops yield critical current densities in the range of 10$^5$-10$^6$~A/cm$^2$, while the corresponding flux-pinning force density reaches several GNm$^{-3}$, underscoring their viability in technological applications. All alloys crystallized in the $\alpha$-W bcc structure and show type-II strongly coupled superconductivity with the isotropic gap. The results obtained from this study on alloys, combined with a malleable morphology required for surface preparation, highlight their potential for further investigation in a thin film form for future applications in superconducting devices, such as single-photon detectors and superconducting qubits.
\begin{figure*} [t]
\includegraphics[width=1.95\columnwidth,origin=b]{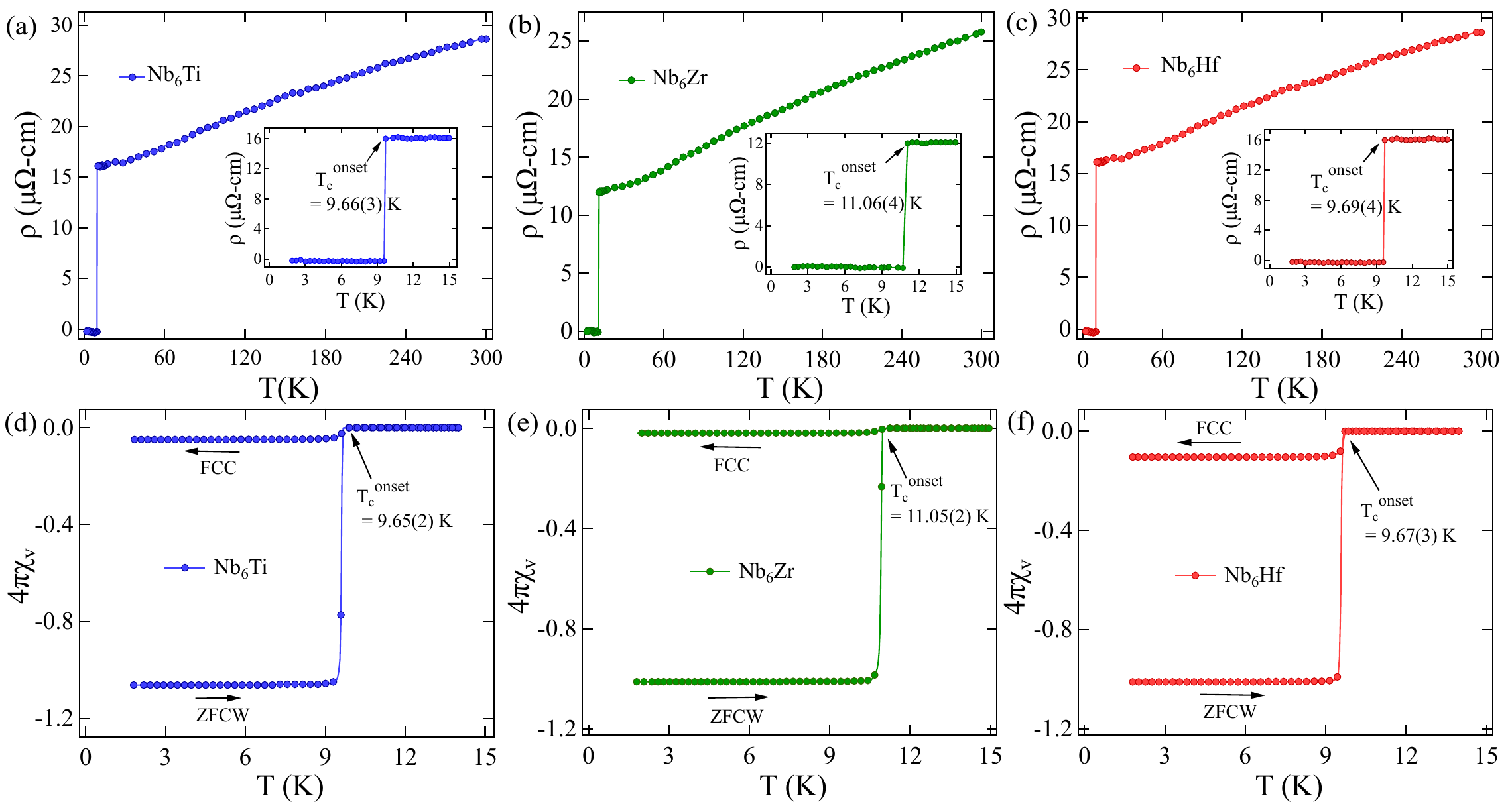}
\caption{\label{RTMT} Temperature-dependent electrical resistivity and the insets show zero drops in resistivity for (a) Nb$_{6}$Ti, (b) Nb$_{6}$Zr, (c) Nb$_{6}$Hf. Magnetization (corrected with demagnetization factor) in ZFCW and FCC mode at an applied field of 1~mT for (d) Nb$_{6}$Ti, (e) Nb$_{6}$Zr, and (f) Nb$_{6}$Hf, respectively.}
\end{figure*}
\section{Experimental Details}
Polycrystalline samples of Nb$_{0.85}$X$_{0.15}$ (X = Ti, Zr, Hf) series were synthesized by arc melting in a pure argon atmosphere on a water-cooled copper hearth. The chamber was first evacuated to a base pressure and then backfilled to a partial pressure. This evacuation and backfilling cycle was performed three times to ensure a highly inert atmosphere. To ensure an oxygen-free environment, a $\text{Zr}$ getter was used to absorb residual gaseous impurities. Stoichiometric amounts of the high-purity constituent elements (purity $\geq 99.9\%$) were melted together on a water-cooled copper hearth using a tungsten electrode. To achieve optimal homogeneity, the resulting ingot/button was flipped and remelted at least four to six times. The buttons were then cooled naturally on the copper hearth. The weight loss of the buttons after melting was observed to be negligible. The final samples used for various measurements were prepared by mechanically sectioning the arc-melted buttons using a low-speed diamond saw with a cutting fluid to prevent localized heating and structural damage, yielding pieces of the required dimensions. Phase purity and crystal structure were analyzed by room-temperature powder X-ray diffraction (XRD) using K$_{\alpha}$ ($\lambda$ = 1.5406~\text {\AA}) radiation using a PANalytical powder X-ray diffractometer. XRD refinement was performed using Fullprof Suite software~\cite{Fullprof}. Elemental composition and homogeneity were assessed by energy-dispersive X-ray analysis (EDXA) on a scanning electron microscope (SEM). Magnetization and AC susceptibility were measured using the Magnetic Property Measurement System (MPMS3). Full range magnetization loops at different temperatures were measured to calculate the critical current density on the rectangular slab of all alloys. Electrical resistivity was measured using the standard four-probe method in 1.9--300~K, while specific heat measurement was performed using the two-tau relaxation technique using the Physical Property Measurement System (PPMS).

\section{Results and Discussion}

\subsection{Sample Characterization}
All alloys are crystallized in the cubic structure $\alpha$ -W, and the crystal structure of Nb$_{0.85}$X$_{0.15}$ (X = Ti, Zr, Hf) is shown in Fig.\ref{XRD}(a). The Rietveld refinement of the Nb$_{0.85}$Ti$_{0.15}$ alloy is shown in Fig.\ref{XRD}(b). The elemental mapping of all alloys is shown in Fig.\ref{XRD}(c). The XRD patterns of the polycrystalline samples of Nb$_{0.85}$X$_{0.15}$, (X = Ti, Zr, Hf) are shown in Fig.\ref{XRD} (d) and the lattice parameters determined from the refinement are summarized in Table\ref{tab1}.

\begin{table}[h]
\small
\caption{Lattice parameters for Nb$_{0.85}$X$_{0.15}$, (X = Ti, Zr, Hf) obtained from XRD refinement.}
\label{tab1}
\begin{center}
\begingroup
\setlength{\tabcolsep}{5pt}
\begin{tabular}[t]{c c c c}\hline
Parameters&Nb$_{0.85}$Ti$_{0.15}$&Nb$_{0.85}$Zr$_{0.15}$&Nb$_{0.85}$Hf$_{0.15}$\\
\hline                                  
a = b = c (\AA)& 3.306(8)& 3.350(1)& 3.349(5)\\
V$_{cell}$ (\AA$^{3}$)& 36.16(6)& 37.59(3) & 37.56(4)\\
\hline
\end{tabular}
\par\medskip\footnotesize
\endgroup
\end{center}
\end{table}

\subsection{Electrical Resistivity}
The temperature-dependent resistivity ($\rho(T)$) of Nb$_6$X (X = Ti, Zr, Hf) measured in the zero field confirms the transition temperatures $T_{c}$ of 9.68(2), 11.06(4), 9.78(3)~K, as shown in Fig.\ref{RTMT}(a), (b), and (c), and the insets show the corresponding zero drops in resistivity, respectively. The low residual resistivity ratio (RRR) is approximately 2, consistent with RRR for polycrystalline samples of some Heusler-type superconducting alloys~\cite{PhysRevB.85.174505}.

\begin{figure*} [t]
\includegraphics[width=1.95\columnwidth,origin=b]{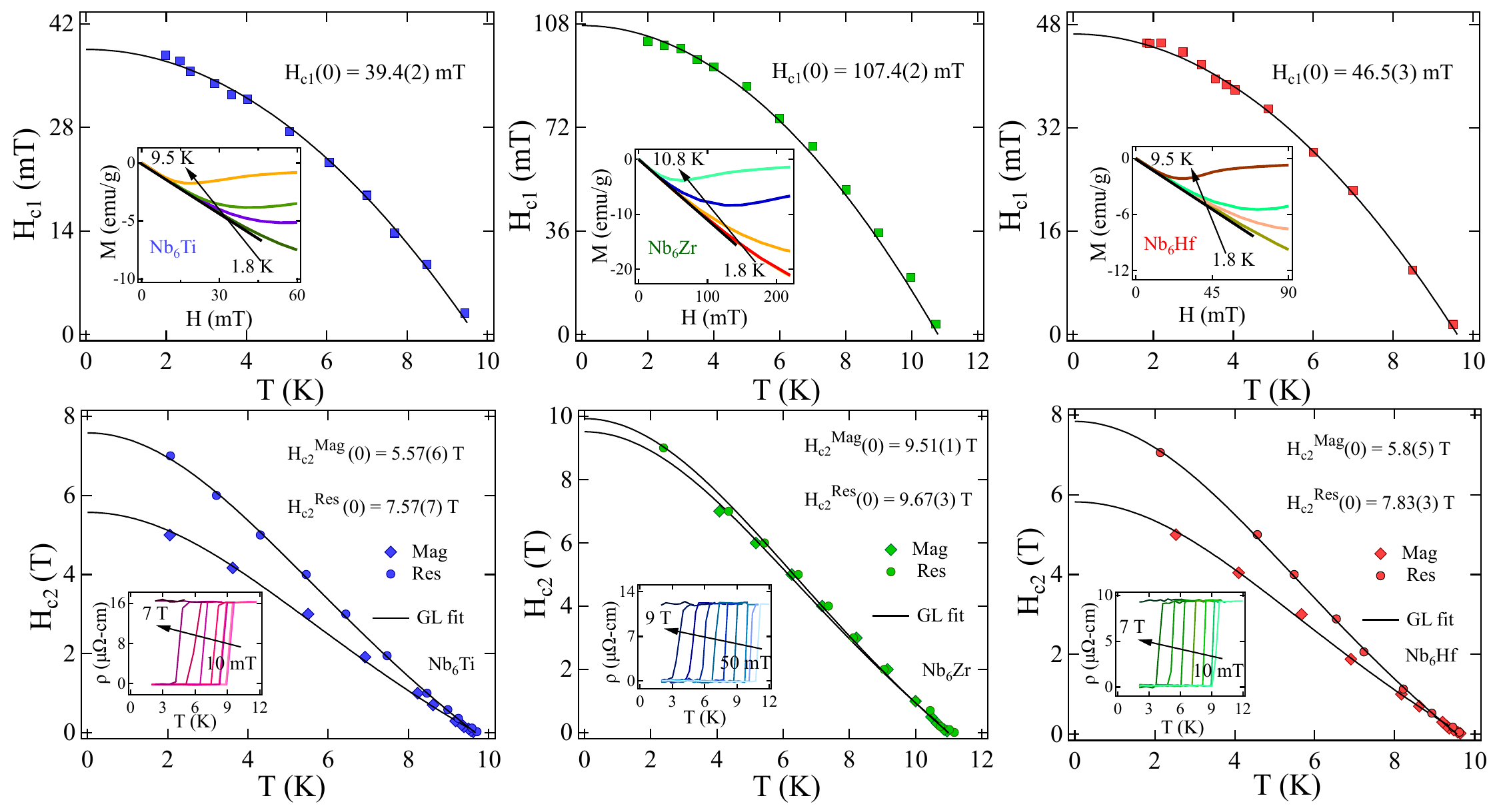}
\caption{\label{Hc1Hc2}Temperature-dependent lower critical field where the solid black curves represent G-L fit using Equation\ref{Hc1} and the insets show the field-dependent magnetization curves for (a) Nb$_{6}$Ti, (b) Nb$_{6}$Zr, and (c) Nb$_{6}$Hf, respectively. Temperature-dependent profiles of the upper critical field, estimated from magnetization and resistivity, where the insets show magnetic field-dependent resistivity data for (d) Nb$_{6}$Ti, (e) Nb$_{6}$Zr, and (f) Nb$_{6}$Hf, respectively.}
\end{figure*}

\subsection{Magnetization}
The superconducting transition temperature of all alloys was confirmed through temperature-dependent magnetization measurements performed in zero field-cooled warming (ZFCW) and field-cooled cooling (FCC) modes under an applied magnetic field of 1~mT, where the quantity $4\pi\chi_{v}$ is corrected for the demagnetization factor. The results, presented in Fig.\ref{RTMT}(d), (e) and (f), indicate $T_c$ values of 9.65 (2) K, 11.05 (2) K and 9.67 (3) K for Nb$_6$X (X = Ti, Zr, Hf), respectively. Temperature and field-dependent magnetization measurements were performed to determine the lower critical field ($H_{c1}(0)$) and the upper critical field ($H_{c2}(0)$). The $H_{c1}$ values at each temperature were identified as the point where the magnetization curve deviates from the Meissner line (solid black line). The temperature dependence of $H_{c1}$ was analyzed using the Ginzburg-Landau (GL) relation given
\begin{equation}
\label{Hc1}
H_{c1}(T) = H_{c1}(0)\left[1-\left(\frac{T}{T_c}\right)^{2}\right]
\end{equation}
which yields $H_{c1}(0)$ = 39.4 (2) mT, 107.4 (2) mT and 46.5 (3) mT for Nb$_6$X (X = Ti, Zr, Hf), respectively. The temperature dependence of $H_{c1}$ is presented in Fig.\ref{Hc1Hc2}(a), (b), and (c), where the insets display magnetization magnetic field curves $vs.$ at various temperatures for each sample. In addition, we estimate $H_{c2}(0)$ through magnetic field and temperature-dependent magnetization and resistivity measurements. The temperature dependence of $H_{c2}$ was plotted and well fitted by the GL equation\ref{Hc2}, as shown in Fig.\ref{Hc1Hc2}(d), (e) and (f) for Nb$_{6}$X, (X = Ti, Zr, Hf), respectively.
\begin{equation}
\label{Hc2}
H_{c2}(T) = H_{c2}(0)\left[\frac{1-(T/T_{c})^{2}}{1+(T/T_{c})^{2}}\right]
\end{equation}
Equation\ref{Hc2} yields estimated values of $H_{c2}(0)$ = 5.57(6), 9.51(1), and 5.81(5)~T from magnetization and 7.57(7), 9.67(3) and 7.83(3)~T from resistivity measurements for Nb$_{6}$X, (X = Ti, Zr, Hf), respectively, much higher than bulk Nb. The discrepancy between the values $H_{c2}(0)$ determined from the resistivity and magnetization measurements likely arises from the inherent limitations of the techniques, as well as their sensitivity to phenomena such as surface superconductivity and flux creep; however, for practical applications, the magnetization-derived value is more reliable, as it better reflects the bulk response of the sample.
\begin{figure*} [t]
\centering
\includegraphics[width=1.9\columnwidth,origin=b]{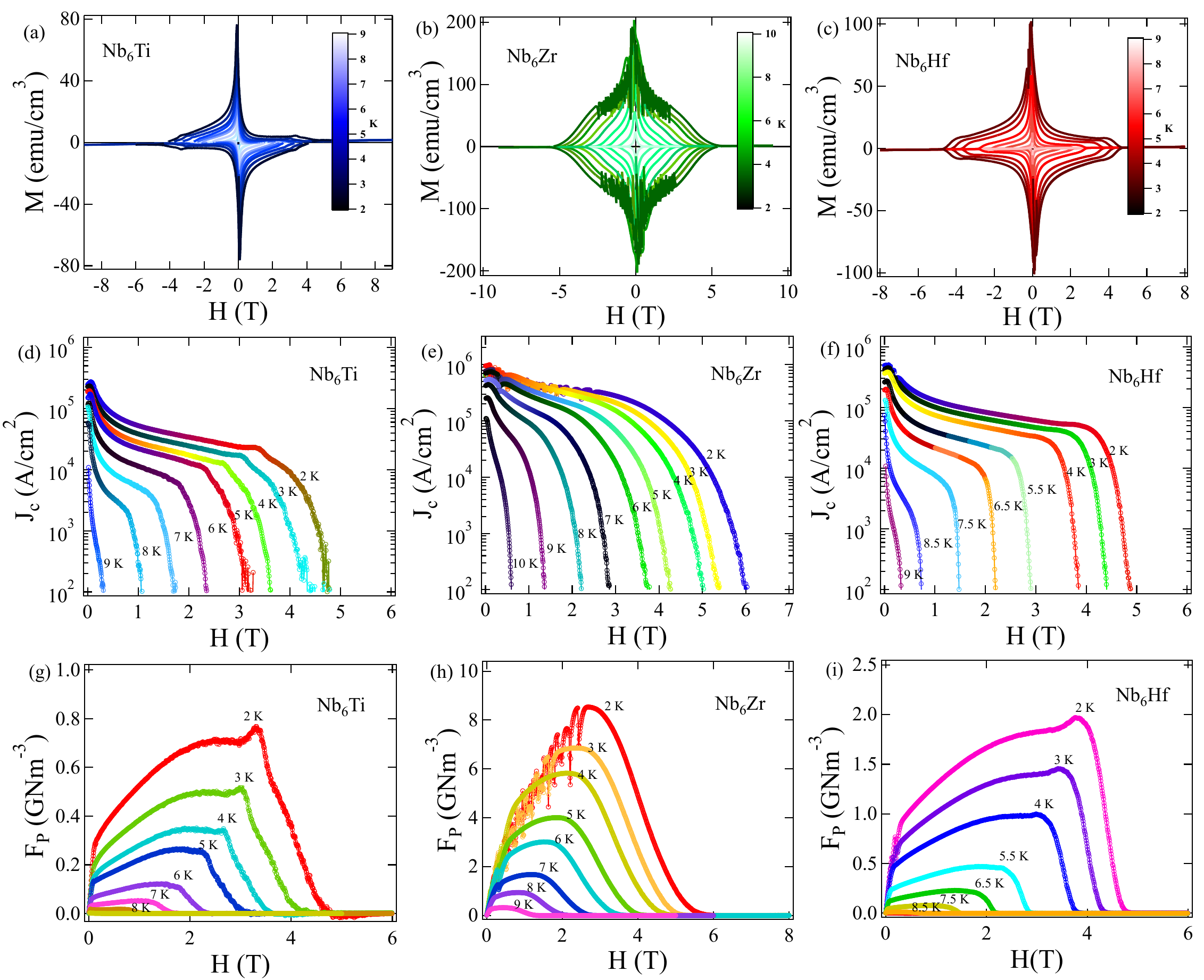}
\caption{\label{Jc} (a), (b) and (c) M-H loops at different temperatures, whereas (d), (e) and (f) magnetic field dependent J$_c$ variation at different temperatures and (g), (h) and (i) Variation of flux pinning force with magnetic field for Nb$_{6}$X, (X = Ti, Zr, Hf), respectively.}
\end{figure*}
The mechanisms behind Cooper pair-breaking when an external magnetic field is applied include the orbital pair-breaking effect and the Pauli limiting field effect. The orbital limiting field, $H_{c2}^{orb}(0)$, can be described by the Werthamer-Helfand-Hohenberg (WHH) model~\cite{WHH1, WHH2}, which neglects the spin-orbit interaction and Pauli paramagnetism. According to the WHH model, $H_{c2}^{orb}(0)$ is expressed as
\begin{equation}
H_{c2}^{orb}(0) = - 0.693~T_{c} \left.\frac{dH_{c2}(T)}{dT}\right\vert_{T=T_{c}} . 
\label{WHH}
\end{equation}
By estimating the initial slopes of the $H_{c2}$ vs $T$ curves, $\frac{-dH_{c2}(T)}{dT}$ at $T = T_c$, we obtain the orbital limiting fields using Equation \ref{WHH} as $H_{c2}^{orb}(0)$ = 4.13(6), 7.11(5), and 4.43(3)~T for Nb$_6$X (X = Ti, Zr, Hf), respectively. For conventional superconductors, the Pauli limit field is given by the expression $H_{c2}^{P} = 1.86 T_c$~\cite{Pauli1, Pauli2}. Using the $T_c$ values obtained from magnetization measurements, we estimate the Pauli limiting fields $H_{c2}^{P}$ as 17.94(3), 20.5(5), and 18.54(2)~T for Nb$_6$X (X = Ti, Zr, Hf), respectively, which are significantly higher than the corresponding estimated values $H_{c2}(0)$. The coherence length $\xi_{GL}(0)$~\cite{Tinkham} can be estimated from $H_{c2}(0)$ using the relation $H_{c2}(0)=\frac{\Phi_{0}}{2\pi\xi_{GL}^{2}}$ where the magnetic flux quantum $\Phi_{0}$ = 2.07 $\times$10$^{-15}$ T m$^{2}$~\cite{Tinkham}. The penetration depth $\lambda_{GL}(0)$~\cite{lambda} is obtained from equation $H_{c1}(0)=\frac{\Phi_{0}}{4\pi\lambda_{GL}^2(0)}\left( ln \frac{\lambda_{GL}(0)}{\xi_{GL}(0)} + 0.12\right)$. The estimated values of $H_{c1}(0)$ and $H_{c2}(0)$ yield coherence lengths $\xi_{GL}(0)$ and penetration depths $\lambda_{GL}(0)$ as 76.9(1), 58.8(5) and 75.6(4)~$\text{\AA}$ for $\xi_{GL}(0)$ and 1073(3), 615(2) and 975(4)~$\text{\AA}$ for $\lambda_{GL}(0)$, for Nb$_{6}$X (X = Ti, Zr, Hf), respectively. The GL parameter defined as $k_{GL}$ = $\frac{\lambda_{GL}(0)}{\xi_{GL}(0)}$, was found to be much higher than $\frac{1}{\sqrt{2}}$ for the three alloys, indicating strong type II superconductivity. In addition, the thermodynamic critical field ($H_{c}(0)$) is defined as $H_{c1}(0)~H_{c2}(0) = H^{2}_{c}(0)~ln(k_{GL})$. This yields the values of $H_{c}(0)$ = 288.6(2), 659.7(5), 325.6(8)~mT for Nb$_{6}$X, (X = Ti, Zr, Hf), respectively. However, RF performance in SRF cavities using bulk Nb reaches the theoretical limit of $H_{c}(0) \sim 200~mT$, our results indicate that the Nb$_{6}$X series has a higher thermodynamic critical field (> 200~mT), hence the higher superheating field ($H_{sh} \approx \frac{0.89}{\sqrt{k_{GL}}}H_c$)~\cite{gurevich2006enhancement}. 

The Ginzburg-Levanyuk number ($Gi$) quantifies the strength of thermal fluctuations relative to vortex unpinning in type-II superconductors~\cite{Gi}, which is described by the following expression
\begin{equation}
Gi= \frac{1}{2} \left(\frac{k_{B} \mu_{0} \tau T_{c}}{4\pi \xi(0)^{3} H_{c}^{2}(0)}\right)^{2} ,
\end{equation}
where $\tau$ is the anisotropy factor that is 1 for cubic Nb$_{6}$X, (X = Ti, Zr, Hf)). Using the estimated $T_{c}$, $\xi(0)$ and $H_{c}(0)$ of magnetization measurement, we obtain $Gi$ = 6.18(2) $\times$ 10$^{-8}$, 1.48(4) $\times$ 10$^{-8}$, 4.43(8) $\times$ 10$^{-8}$ for Nb$_{6}$X, (X = Ti, Zr, Hf), respectively. These values are comparable to typical values $Gi$ observed for low $T_{c}$ superconductors ($\sim$10$^{-8}$). This suggests that thermal fluctuations do not contribute significantly to vortex unpinning in Nb$_{6}$X series~\cite{lowTcGi}.

\begin{figure*} [t]
\centering
\includegraphics[width=1.95\columnwidth,origin=b]{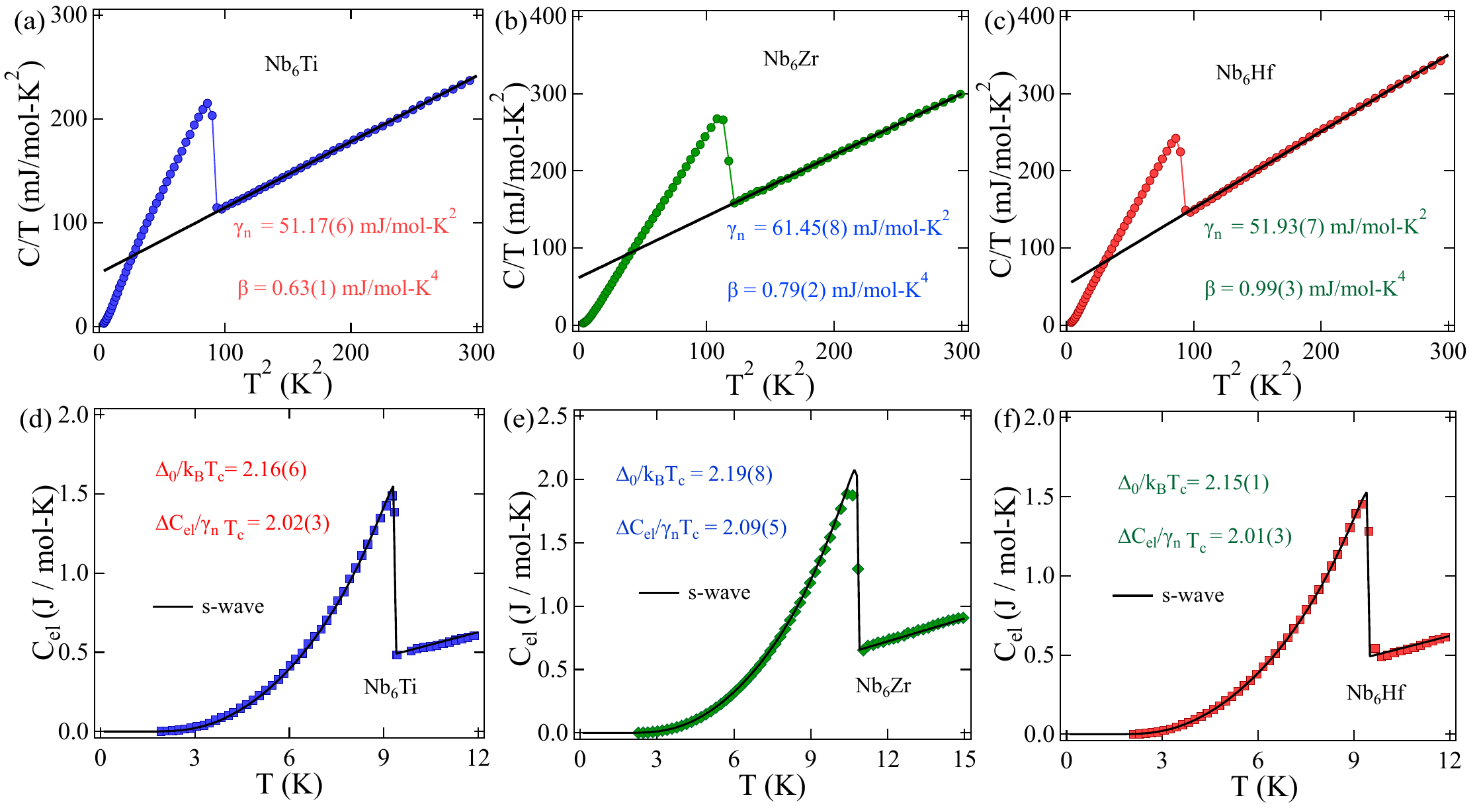}
\caption{\label{SH} (a), (b) and (c) C/T vs T$^{2}$ plot for Nb$_{6}$X, (X = Ti, Zr, Hf), respectively, where the solid black lines represent the Debye-Sommerfeld fitting represented by Equation\ref{C/T} (d), (e) and (f) C$_{el}$ vs T plot for Nb$_{6}$X, (X = Ti, Zr, Hf), respectively, where solid black curves represent the s-wave model fit with BCS type single gap function.}
\end{figure*}

\subsection{Critical Current Density and Flux Pinning Force}
Magnetic hysteresis loops (see Figs.\ref{Jc}(a), (b), and (c)) of Nb$_{6}$X (X = Ti, Zr, Hf) series were measured at different temperatures to assess the critical current density ($J_c$) and its variation with magnetic field. Both Nb$_{6}$Zr and Nb$_{6}$Hf alloys show several low-field flux jumps in the M-H curves, as shown in Fig.\ref{Jc}(b) and (c). These flux jumps weaken and disappear when the temperature rises above 4~K or the magnetic field exceeds 3~T. Usually, flux jump phenomena are common in these alloys because of the strong magnetic flux pinning, while Nb$_{6}$ Ti does not have a flux pinning force at all, suggesting a relatively weak flux pinning force. The magnetic flux pinning force inside the Nb$_{6}$Zr sample is strongest, reflecting the high-low field flux jumps. The irreversible field (H$_{irr}$) of all samples is evaluated from the corresponding M-H loops at different temperatures. H$_{irr}$ is the magnetic field above which vortice unpinning begins, and as the temperature increases, the irreversible fields of all samples decrease to various degrees. $J_c$ is obtained by the Bean model~\cite{bean1964magnetization} as described for $J_c$ in the following equation:
\begin{equation}
J_c = \frac{20 ~\Delta M}{w \left(1- \frac{w}{3l}\right)}
\label{Jc}
\end{equation}
where, w and l are the width and length of the sample (l $\gg$ w) perpendicular to the direction of applied magnetic field, and $\Delta M$ is the width of the magnetization at the same magnetic field and Fig.\ref{Jc} (d), (e) and (f) show the magnetic field dependence of $J_c$ for Nb$_{6}$X, (X = Ti, Zr, Hf), respectively. The $J_{c}(0)$ at T = 2 K is extracted as $J_{c}(0)$ = 2.48 $\times$ 10$^5$ A/cm$^2$, 1.02 $\times$ 10$^6$~A/cm$^2$ and 4.87 $\times$ 10$^5$ A/cm$^2$ for Nb$_{6}$ X (X = Ti, Zr, Hf), respectively. In addition, the pinning force density was calculated using the relation $F_p = \mu_0 H \times J_c$~\cite{PhysRevB.75.134515}. The variations of the flux pinning force density with magnetic field for Nb$_{6}$X, (X = Ti, Zr, Hf) are shown in Fig.\ref{Jc}(g), (h) and (i), respectively. The density of the pinning force is obtained in the order of GNm$^{-3}$ for this series, with the highest value obtained for Nb$_6$Zr due to the presence of low-field flux jumps, which is relatively low in Nb$_6$Hf and negligible in Nb$_6$Ti, suggesting the lowest value of the pinning force in Nb$_6$Ti. While flux jumps are commonly observed in Nb-based superconductors due to thermomagnetic instabilities, their absence in certain Nb alloys can be attributed to enhanced thermal conductivity and homogeneous flux pinning, which suppress abrupt vortex avalanches~\cite{fluxjumpsinNb}. The critical current density and flux pinning force values for all alloys exceed the practical threshold value.

\subsection{Specific Heat}
Temperature-dependent specific heat measurements were performed for all samples without an external magnetic field to study the thermal properties of Nb$_{6}$X (X = Ti, Zr, Hf). Normal state $C/T$ vs $T^{2}$ data were fitted using the Debye-Sommerfeld model represented by Equation\ref{C/T} and shown in Fig.\ref{SH}(a),(b) and (c) with solid black curves.
\begin{equation}
C/T = \gamma_{n}+\beta~T^{2}
\label{C/T}
\end{equation}
where, $\gamma_{n}$ is the Sommerfeld coefficient, $\beta$ holds the phononic contributions to the specific heat. Fitting normal state specific heat data using Equation\ref{C/T} gives $\gamma_{n}$ = 51.17(6), 61.45(8), 51.93(7)~mJ mol$^{-1}$K$^{-2}$ and $\beta$ = 0.63(1), 0.79(2), 0.99(3)~mJ mol$^{-1}$K$^{-4}$ for Nb$_{6}$X (X = Ti, Zr, Hf), respectively. $\gamma_{n}$ is related to the density of states on the Fermi surface $D_{c}(E_{f})$ by the relation $\gamma_{n}$ = $\left(\frac{\pi^{2}k_{B}^{2}}{3}\right)D_{c}(E_{f})$, where $k_{B}$ $\approx$ 1.38 $\times$ 10$^{-23}$~J K$^{-1}$. $D_{c}(E_{f})$ is estimated to be 21.7(1), 26.1(2), 22.1(1)~states/eV f.u. for Nb$_{6}$X, (X = Ti, Zr, Hf), respectively. The debye temperature ($\theta_{D}$) related to $\beta$ as $\theta_{D}$ = $\left(\frac{12\pi^{4}RN}{5\beta_{3}}\right)^{1/3}$, where $N$ is the number of atoms per formula unit and $R$ is the molar gas constant (8.314~J mol$^{-1}$ K$^{-1}$), which calculates $\theta_{D}$ = 277.8(9), 258.2(4), 239.5(3)~K for Nb$_{6}$X, (X = Ti, Zr, Hf), respectively.

McMillan's model estimates the electron-phonon coupling strength from a dimensionless quantity $\lambda_{e-ph}$~\cite{McM}, which depends on the estimated values of $\theta_{D}$ and $T_{c}$ as
\begin{equation}
\lambda_{e-ph} = \left[\frac{1.04 + \mu^{*}ln(\theta_{D}/1.45T_{c})}{(1 - 0.62\mu^{*})ln(\theta_{D}/1.45T_{c}) - 1.04}\right]
\end{equation}
where $\mu^{*}$ is screened Coulomb repulsion, considering $\mu^{*}$ = 0.13 as described for transition metals~\cite{McM}, estimated $\lambda_{e-ph}$ = 0.83(6), 0.92(4), 0.89(7) for Nb$_{6}$X, (X = Ti, Zr, Hf), respectively. It classifies them as strongly coupled superconductors. After subtracting the term $\beta~T^3$ from total specific heat ($C$), the temperature-dependent electronic specific heat ($C_{el}$) was extracted for all samples and fitted with the isotropic single gap model, represented as solid black curves in Fig\ref{SH}(d), (e) and (f) and the quantity $\frac{\Delta C_{el}}{\gamma_{n}T_{c}}$ = 1.87(2), 1.79(5), 1.85(3) for Nb$_{6}$X, (X = Ti, Zr, Hf), respectively.

Further, $C_{el}$ $vs.$ $T$ data can provide information about the symmetry of the superconducting gap around the Fermi surface, revealing the intricate pairing mechanism. Normalized entropy ($S$) in the superconducting region and $C_{el}$ can be related as
\begin{equation}
\frac{C_{el}}{\gamma_{n}T_{c}} = t ~\frac{d\left(S/\gamma_{n}T_{c}\right)}{dt}
\end{equation}
where $t = T/T_{c}$, the reduced temperature. Within the framework of the BCS approximation~\cite{BCS1,BCS2}, the normalized entropy for a single BCS-like gap is defined by the following relation.
\begin{equation}
\label{BCSeq}
\frac{S}{\gamma_{n}T_{c}} = -\frac{6}{\pi^2}\Delta^{s}_{T}\int_{0}^{\infty}[ \textit{f}\ln(f)+(1-f)\ln(1-f)]dy
\end{equation}
where, $\Delta^{s}_{T} = \frac{\Delta(0)}{k_{B}T_{c}}$, $\textit{f}$($\xi$) = [exp($\textit{E}$($\xi$)/$k_{B}T$)+1]$^{-1}$ is the Fermi function, $\textit{E}$($\xi$) = $\sqrt{\xi^{2}+\Delta^{2}(t)}$, where $E(\xi)$ is the energy of normal electrons measured relative to the Fermi energy, $\textit{y}$ = $\xi/\Delta(0)$, and $\Delta(t) = \tanh[1.82(1.018((\mathit{1/t})-1))^{0.51}$] represents the temperature-dependent superconducting energy gap. The solid black curves in Fig.\ref{SH}(d), (e), and (f) represent the fit to electronic specific heat data using a single isotropic nodeless gap model as described by Equation\ref{BCSeq}, which provides the normalized superconducting gap values $\Delta(0)/k_B T_c$ = 2.16(1), 2.19(6), 2.15(3) for Nb$_{6}$X (X = Ti, Zr, Hf), respectively, which confirms the strong electron-phonon coupled superconductivity.

\subsection{Normal State Properties and Uemura Classification}
In this section, we have analytically solved a set of equations simultaneously to investigate electronic properties and discussed the Uemura classification of Nb$_{6}$X (X = Ti, Zr, Hf)~\cite{Uemura}. The Sommerfeld coefficient ($\gamma_{n}$) is related to the quasiparticle number-density ($n$), defined as
\begin{equation}
\gamma_{n}= \left(\frac{\pi}{3}\right)^{2/3} \frac{k_{B}^{2} m^{\ast} V_{f.u.} n^{1/3}}{\hbar^{2} N_{A}}
\end{equation}
 and Fermi velocity ($v_{F}$) is related to the electronic mean free path ($l_{e}$) and $n$ by the expressions 
 \begin{equation}
 l_{e} = \frac{3 \pi^{2} \hbar^{2}}{e^{2} \rho_{0} m\ast^{2} v_{F}^2} ~\text{and}~  
 n=\frac{1}{3 \pi^{2}} \left(\frac{m^{\ast} v_{F}}{\hbar}\right)^{3}
 \end{equation}
respectively, where $k_{B}$ is the Boltzmann constant, $m^{\ast}$ is the effective mass of the quasiparticles, $V_{f.u.}$ is the volume of the formula unit, $N_{A}$ is the Avogadro number and $\rho_{0}$ is the residual resistivity.
 
\begin{figure} [t]
\centering
\includegraphics[width=0.95\columnwidth,origin=b]{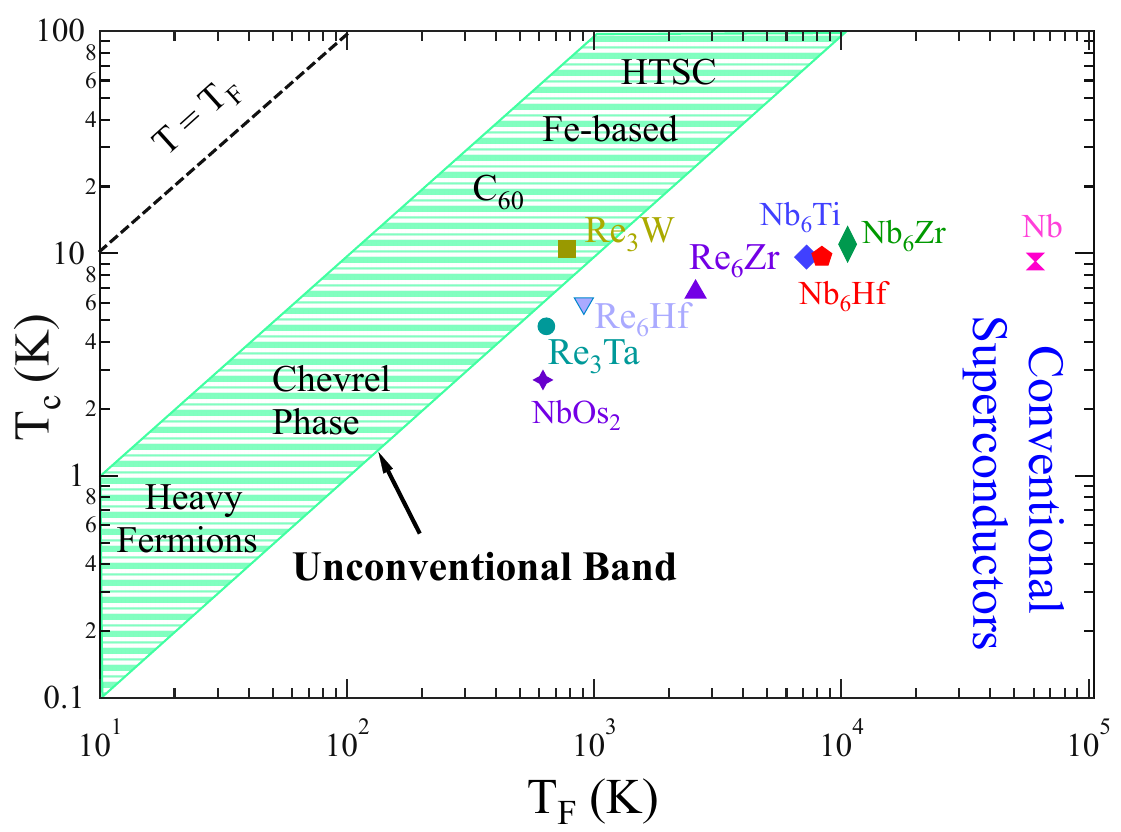}
\caption{\label{Uemura} Uemura plot defined by $T_{c}$ and $T_{F}$ for Nb$_{6}$X (X = Ti, Zr, Hf), where the green shaded region represents the unconventional band.}
\end{figure}
\begin{figure*} [t]
\centering
\includegraphics[width=1.25\columnwidth,origin=b]{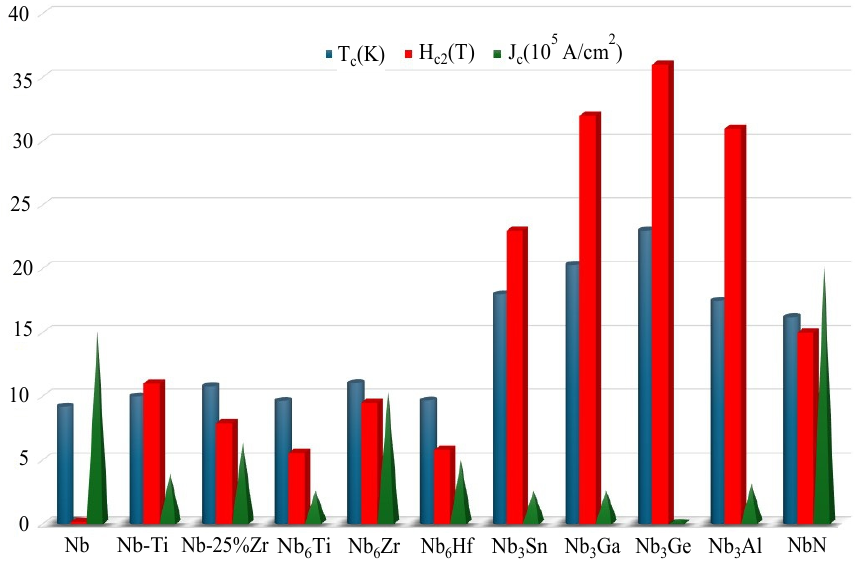}
\caption{\label{Bar_Nb} Comparison of T$_c$, H$_{c2}$, J$_{c}$ of Nb$_{6}$X, (X = Ti, Zr, Hf) with niobium\cite{bean1964magnetization} and its alloys~\cite{Xu2021Nb3Sn,NbZrPhysRevLett.6.671,NbZrPRM,Nb0.75Zr0.25,fietz1964magnetization,mirmefstein1997mixed,junod2002specificNb0.77Zr0.23,Nb3Al_recent,Thompson1979Nb3Ge,Webb1973Nb3Ge,Flukiger1981Nb3Ga,Joshi2018NbN,Jing2023NbN,AIP2023NbN}.}
\end{figure*}
Within the dirty limit superconductivity, the BCS coherence length of a superconductor is much larger than the mean free path ($\frac{\xi_{0}}{l_{e}} \gg 1$), and the scattering of electrons with impurities and defects may affect the superconducting properties. The GL penetration depth ($\lambda_{GL}$) is related to the London penetration depth ($\lambda_{L}$) at absolute zero temperature followed by the expression
\begin{equation}
\lambda_{GL}(0)=\lambda_{L}\left(1+\frac{\xi_{0}}{l_{e}}\right)^{1/2}, \text{where} 
~\lambda_{L}=\left[\frac{m^{\ast}}{\mu_{0}ne^{2}}\right]^{1/2} 
\end{equation}
and BCS coherence length relates to GL coherence length and is expressed by
\begin{equation}
\frac{\xi_{GL}(0)}{\xi_{0}}= \frac{\pi}{2\sqrt{3}}(1+\frac{\xi_{0}}{l_{e}})^{-1/2}. 
\end{equation}
We have solved the above equations simultaneously and estimated the values of $m^{\ast}$, $n$, $v_{F}$, $\xi_{0}$, $l_{e}$, using the previously obtained values of $\gamma_{n}$, $\rho_{0}$, $\xi_{GL}(0)$ and $\lambda_{GL}(0)$. The Fermi temperature for an isotropic spherical Fermi surface is defined as
\begin{equation}
T_{F} = \frac{\hbar^{2}k_{F}^{2/3}} {2~m^{\ast} k_{B}}\quad, 
\end{equation}

where $k_{F}=3\pi^{2}n$, the Fermi wave vector. The ratio $T_{c}/T_{F}$ classifies superconductors into the conventional or unconventional category. According to Uemura et al.~\cite{Uemura}, the unconventional range 0.01$ \leq$ $\frac{T_{c}}{T_{F}}$ $\leq$ 0.1,  shown by the green shaded band in Fig.\ref{Uemura}, whereas, $T_{c}/T_{F}$ values for Nb$_{6}$X (X = Ti, Zr, Hf), lie outside this unconventional region. All superconducting and normal state parameters of Nb$_{6}$X (X = Ti, Zr, Hf), estimated from various techniques, are listed in Table \ref{tbl2}. This comparison highlights that, despite variations in microstructure and alloying, Nb-based systems generally exhibit weak to moderate electronic correlations and conventional electron-phonon-mediated pairing. The similarity of $T_{c}/T_{F}$ values across these families underscores that superconductivity in Nb$_{6}$X alloys is governed by the same fundamental mechanism, situating them firmly within the conventional regime while still offering competitive superconducting parameters for practical applications.
\begin{table} [t]
\small
\caption{Superconducting and normal state parameters of Nb$_{6}$X, (X = Ti, Zr, Hf).}
\label{tbl2}
\begin{center}
\begin{tabular}
{p{0.2\linewidth}p{0.16\linewidth}p{0.16\linewidth}p{0.16\linewidth}p{0.12\linewidth}}
\hline
Parameters & unit &Nb$_{6}$Ti &Nb$_{6}$Zr &Nb$_{6}$Hf\\
\hline
$T_{c}$& K& 9.65(2)& 11.05(2)& 9.67(3)\\           
$H_{c1}(0)$& mT& 39.4(2)& 107.4(2)& 46.5(3)\\
$H_{c2}(0)$& T& 5.57(6)& 9.51(1) & 5.81(5)\\
$H_{c2}(0)^{Pauli}$& T& 17.94(3)& 20.5(5)& 18.54(2)\\
$H_{c2}(0)^{orb}$& T& 4.13(6)& 7.11(5)& 4.43(3)\\
$\xi_{GL}(0)$&  \text{\AA}& 76.9(1)& 58.8(5)& 75.6(4)\\
$\lambda_{GL}(0)$& \text{\AA}& 1073(3)& 615(2) &975(4)\\
$H_{c}(0)$& mT& 288.6(2) &659.7(5) &325.6(8) \\
$k_{GL}$& &13.9(5) &10.4(5) &12.9(4)\\
$\Delta C_{el}/\gamma_{n}T_{c}$& &2.02(3) &2.09(5) &2.01(3)\\
$\Delta(0)/k_{B}T_{c}$&  &2.16(6)&2.19(8)& 2.15(1)\\
$\theta_{D}$& K& 277.8(9)& 258.2(4)& 239.5(3)\\
$\lambda_{e-ph}$& & 0.83(6)& 0.92(4)& 0.89(7)\\
$v_{F}$&$10^{5}$ m$s^{-1}$& 1.2(4)&1.4(6)& 1.3(2)\\
$n$&$10^{28}$ $m^{-3}$& 13.4(2)& 23.5(4)& 15.6(3)\\
$T_{F}$&K& 7234(4)& 10595(25)& 8346(8)\\
${T_{c}}/{T_{F}}$& & 0.0013(3)& 0.0010(3)&0.0011(5)\\
${m^{*}}/{m_{e}}$& & 15.2(3)& 15.1(2)& 14.6(2)\\
\hline
\end{tabular}
\end{center}
\par\medskip\footnotesize
\end{table}

\section{Discussion}
Figure~\ref{Bar_Nb} presents a comparison of key superconducting parameters ($T_c$, $H_{c2}$ and $J_c$) for Nb--T alloys $\alpha$--W type (T = Ti, Zr, Hf) and other well-known compounds based on niobium. The data provide insight into the variation in superconducting behavior between different material classes. Niobium-based superconductors, especially binary alloys of Nb-T (where T = Ti, Zr, Hf) and Nb$_3$X compounds of type A15, represent two technologically significant families of low-temperature superconductors, each offering unique advantages tailored to specific operational demands. Nb-Ti alloys are widely used in commercial superconducting technologies due to their good thermal performance and suitability for wire manufacturing on a large scale~\cite{NbTihon2003composition,NbTiPhysRevB.110.L140502,NbTipressure,TixNb}. Introducing alloying elements such as Ti or Zr into niobium enhances flux pinning capabilities while maintaining critical superconducting parameters, with transition temperatures ranging from 9 to 11 K and upper critical fields in the range of 6--14~T~\cite{Shapira1965} at ambient pressure. Specifically, incorporating zirconium into niobium has been shown to enhance surface superconductivity and radio-frequency characteristics, aided by the formation of a stable ZrO$_2$ surface layer and improved electron-phonon interactions~\cite{ZrdopedNb}. In addition, alloys such as Nb-Zr and Nb--Hf exhibit strong resistance to radiation damage, making them particularly suitable for nuclear environments~\cite{He2020}. Recent computational studies suggest that irradiation promotes the formation of Nb-rich precipitates within Zr matrices, which adopt energetically favorable platelet morphologies. These structures contribute to enhanced microstructural integrity in nuclear fuel cladding materials under irradiation conditions~\cite{Goel2021,Azevedo2011}.

In contrast, A15-type intermetallics such as Nb$_3$Sn, Nb$_3$Al, and Nb$_3$Ge demonstrate significantly enhanced superconducting characteristics, with critical temperatures ($T_c$) reaching up to 23~K and upper critical magnetic fields ($H_{c2}$) exceeding 30~T. These remarkable properties are mainly attributed to their distinctive crystal structure, characterized by quasi-one-dimensional chains of Nb atoms and a densely packed A$_3$B lattice~\cite{NTSS}. This arrangement strengthens electron-phonon interactions near the Fermi level, thereby promoting superconductivity. However, these materials are inherently brittle and exhibit a high sensitivity to slight variations in composition, which presents challenges during processing. Beyond the superconducting parameters summarized in Fig\ref{Bar_Nb}, prior studies have shown that microstructural refinement (e.g., grain size control in Nb$_3$Sn~\cite{Xu2021Nb3Sn} and stacking fault regulation by nano-oxide particles in Nb$_3$Al~\cite{Nb3Al_recent}), defect density introduced by irradiation~\cite{Eisterer2018Irradiation}, and processing history such as hot rolling or equal channel angular pressing (ECAP) in NbTi~\cite{Zhao2021NbTi} critically determine flux pinning and current-carrying capacity. These insights also contextualize the potential of Nb$_6$X alloys. The bulk properties observed here motivate future efforts in thin-film growth and characterization, where enhanced superconducting performance and device integration could be realized.

\section{Conclusion}
In summary, we have synthesized and characterized the Nb-rich series, Nb$_{0.85}$X$_{0.15}$ (nominal composition: Nb$_6$X, X = Ti, Zr, Hf), which crystallizes in a $\alpha$-W bcc structure. We have thoroughly investigated their superconducting and normal state properties through XRD, electrical resistivity, magnetization, and specific heat measurements. All normal and superconducting state parameters follow nonmonotonic trends in Nb$_{6}$X series, since the X component is substituted by elements of Ti (3d), Zr (4d), and Hf (5d). The calculated critical current density from temperature-dependent magnetization loops has values in the range 10$^5$--10$^6$~A/cm$^2$ with a flux pinning force of the order of GNm$^{-3}$, which are good values for practical applications.  All alloys have moderate transition temperatures and upper critical fields following isotropic nodeless superconducting gaps with strong electron-phonon coupling strength. The variation in spin-orbit coupling strength from 3d to 5d elements, while maintaining a fixed Nb content, shows only a negligible influence on superconducting properties and gap symmetry, but it can provide further insight into the superconducting ground state and its relation to mechanical properties. Due to their morphological malleability and superior superconducting properties, this family of binary alloys holds significant promise, and dedicated thin-film studies will be essential to fully explore their technological potential in superconducting devices.

\section{Acknowledgement}
R~P~S acknowledges the SERB Government of India for the Core Research Grant No. CRG/2023/000817. R~K~K acknowledges the UGC, Government of India, for providing a Senior Research Fellowship.
 
\bibliography{references}

@article{NTSS,
  title = {Topological states in \text{A15} superconductors},
  author = {Kim, Minsung and Wang, Cai-Zhuang and Ho, Kai-Ming},
  journal = {Phys. Rev. B},
  volume = {99},
  issue = {22},
  pages = {224506},
  numpages = {5},
  year = {2019},
  month = {Jun},
  publisher = {American Physical Society},
  doi = {10.1103/PhysRevB.99.224506},
  url = {https://link.aps.org/doi/10.1103/PhysRevB.99.224506}
}

@article{LargeMRNb3Sb,
year = {2021},
month = {sep},
publisher = {Chinese Physical Society and IOP Publishing Ltd},
volume = {38},
number = {8},
pages = {087501},
author = {Qin Chen and Yuxing Zhou and Binjie Xu and Zhefeng Lou and Huancheng Chen and Shuijin Chen and Chunxiang Wu and Jianhua Du and Hangdong Wang and Jinhu Yang and Minghu Fang},
title = {Large Magnetoresistance and Nontrivial Berry Phase in \text{Nb$_{3}$Sb} Crystals with \text{A15} Structure},
journal = {Chinese Physics Letters},
doi = {10.1088/0256-307X/38/8/087501},
url = {https://dx.doi.org/10.1088/0256-307X/38/8/087501}
}

@article{NbRF1,
author = {Proch, Dieter and Schmueser, Peter and Singer, Waldemar and Lilje, L.},
year = {2001},
month = {01},
pages = {},
title = {Niobium in superconducting \text{RF} cavities},
journal = {Niobium, Science and Technology},
url = {https://www.researchgate.net/publication/237733978_Niobium_in_superconducting_RF_cavities}
}

@article{NbRFsurface,
  title={Surface superconductivity in niobium for superconducting RF cavities},
  author={Casalbuoni, S and Knabbe, EcA and K{\"o}tzler, J and Lilje, L and Von Sawilski, L and Schmueser, Peter and Steffen, B},
  journal={Nucl. Instrum. Methods Phys. Res. A},
  volume={538},
  number={1-3},
  pages={45--64},
  year={2005},
  publisher={Elsevier},
  url = {https://doi.org/10.1016/j.nima.2004.09.003}
}

@article{Nbcavity,
  title={Understanding mechanism of performance improvement in nitrogen-doped niobium superconducting radio frequency cavity},
  author={Fang, Xiaotian and Oh, Jin-Su and Kramer, Matt and Romanenko, A and Grassellino, A and Zasadzinski, John and Zhou, Lin},
  journal={Materials Research Letters},
  volume={11},
  number={2},
  pages={108--116},
  year={2023},
  publisher={Taylor \& Francis},
  url = {https://doi.org/10.1080/21663831.2022.2126737},
}

@article{Fullprof,
title = {Recent advances in magnetic structure determination by neutron powder diffraction},
journal = {Physica B: Condensed Matter},
volume = {192},
number = {1},
pages = {55-69},
year = {1993},
issn = {0921-4526},
doi = {https://doi.org/10.1016/0921-4526(93)90108-I},
url = {https://www.sciencedirect.com/science/article/pii/092145269390108I},
author = {Juan Rodríguez-Carvajal}
}

@book{tinkham,
  title={Introduction to superconductivity},
  author={Tinkham, Michael},
  volume={1},
  year={2004},
  publisher={Courier Corporation}
}

@article{WHH1,
  title = {Temperature and Purity Dependence of the Superconducting Critical Field, \text{H$_{c2}$. II}},
  author = {Helfand, E. and Werthamer, N. R.},
  journal = {Phys. Rev.},
  volume = {147},
  issue = {1},
  pages = {288--294},
  numpages = {0},
  year = {1966},
  month = {Jul},
  publisher = {American Physical Society},
  doi = {10.1103/PhysRev.147.288},
  url = {https://link.aps.org/doi/10.1103/PhysRev.147.288}
}

@article{WHH2,
  title = {Temperature and Purity Dependence of the Superconducting Critical Field, \text{H$_{c2}$. III}. Electron Spin and Spin-Orbit Effects},
  author = {Werthamer, N. R. and Helfand, E. and Hohenberg, P. C.},
  journal = {Phys. Rev.},
  volume = {147},
  issue = {1},
  pages = {295--302},
  numpages = {0},
  year = {1966},
  month = {Jul},
  publisher = {American Physical Society},
  doi = {10.1103/PhysRev.147.295},
  url = {https://link.aps.org/doi/10.1103/PhysRev.147.295}
}

@article{Pauli1,
  title = {Structure and physical properties of the noncentrosymmetric superconductor \text{Mo$_{3}$Al$_{2}$C}},
  author = {Karki, A. B. and Xiong, Y. M. and Vekhter, I. and Browne, D. and Adams, P. W. and Young, D. P. and Thomas, K. R. and Chan, Julia Y. and Kim, H. and Prozorov, R.},
  journal = {Phys. Rev. B},
  volume = {82},
  issue = {6},
  pages = {064512},
  numpages = {7},
  year = {2010},
  month = {Aug},
  publisher = {American Physical Society},
  doi = {10.1103/PhysRevB.82.064512},
  url = {https://link.aps.org/doi/10.1103/PhysRevB.82.064512}
}

@article{Pauli2,
  title = {Superconductivity in Quasi-One-Dimensional \text{K$_{2}$Cr$_{3}$As$_{3}$} with Significant Electron Correlations},
  author = {Bao, Jin-Ke and Liu, Ji-Yong and Ma, Cong-Wei and Meng, Zhi-Hao and Tang, Zhang-Tu and Sun, Yun-Lei and Zhai, Hui-Fei and Jiang, Hao and Bai, Hua and Feng, Chun-Mu and Xu, Zhu-An and Cao, Guang-Han},
  journal = {Phys. Rev. X},
  volume = {5},
  issue = {1},
  pages = {011013},
  numpages = {6},
  year = {2015},
  month = {Feb},
  publisher = {American Physical Society},
  doi = {10.1103/PhysRevX.5.011013},
  url = {https://link.aps.org/doi/10.1103/PhysRevX.5.011013}
}

@article{lambda,
  title = {Physical Properties of the Noncentrosymmetric Superconductor \text{Mg$_{10}$Ir$_{19}$B$_{16}$}},
  author = {Klimczuk, T. and Ronning, F. and Sidorov, V. and Cava, R. J. and Thompson, J. D.},
  journal = {Phys. Rev. Lett.},
  volume = {99},
  issue = {25},
  pages = {257004},
  numpages = {4},
  year = {2007},
  month = {Dec},
  publisher = {American Physical Society},
  doi = {10.1103/PhysRevLett.99.257004},
  url = {https://link.aps.org/doi/10.1103/PhysRevLett.99.257004}
}

@article{McM,
  title = {Transition Temperature of Strong-Coupled Superconductors},
  author = {McMillan, W. L.},
  journal = {Phys. Rev.},
  volume = {167},
  issue = {2},
  pages = {331--344},
  numpages = {0},
  year = {1968},
  month = {Mar},
  publisher = {American Physical Society},
  doi = {10.1103/PhysRev.167.331},
  url = {https://link.aps.org/doi/10.1103/PhysRev.167.331}
}

@article{BCS1,
  title = {Theory of Superconductivity},
  author = {Bardeen, J. and Cooper, L. N. and Schrieffer, J. R.},
  journal = {Phys. Rev.},
  volume = {108},
  issue = {5},
  pages = {1175--1204},
  numpages = {0},
  year = {1957},
  month = {Dec},
  publisher = {American Physical Society},
  doi = {10.1103/PhysRev.108.1175},
  url = {https://link.aps.org/doi/10.1103/PhysRev.108.1175}
}

@article{BCS2,
  title = {Exchange Scattering in Superconductors},
  author = {Suhl, H. and Matthias, B. T.},
  journal = {Phys. Rev. Lett.},
  volume = {2},
  issue = {1},
  pages = {5--6},
  numpages = {0},
  year = {1959},
  month = {Jan},
  publisher = {American Physical Society},
  doi = {10.1103/PhysRevLett.2.5},
  url = {https://link.aps.org/doi/10.1103/PhysRevLett.2.5}
}

@article{Gi,
title = {Multiband superconductivity in \text{Lu$_{3}$Os$_{4}$Ge$_{13}$}},
doi = {10.1088/0953-2048/28/11/115012},
url = {https://dx.doi.org/10.1088/0953-2048/28/11/115012},
year = {2015},
month = {sep},
publisher = {IOP Publishing},
volume = {28},
number = {11},
pages = {115012},
author = {Om Prakash and A Thamizhavel and S Ramakrishnan},
journal = {Supercond. Sci. Technol.}
}

@article{Uemura,
  title = {Systematic variation of magnetic-field penetration depth in high-\text{T$_c$} superconductors studied by muon-spin relaxation},
  author = {Uemura, Y. J. and Emery, V. J. and Moodenbaugh, A. R. and Suenaga, M. and Johnston, D. C. and Jacobson, A. J. and Lewandowski, J. T. and Brewer, J. H. and Kiefl, R. F. and Kreitzman, S. R. and Luke, G. M. and Riseman, T. and Stronach, C. E. and Kossler, W. J. and Kempton, J. R. and Yu, X. H. and Opie, D. and Schone, H. E.},
  journal = {Phys. Rev. B},
  volume = {38},
  issue = {1},
  pages = {909--912},
  numpages = {0},
  year = {1988},
  month = {Jul},
  publisher = {American Physical Society},
  doi = {10.1103/PhysRevB.38.909},
  url = {https://link.aps.org/doi/10.1103/PhysRevB.38.909}
}

@article{PhysRevMaterials.8.084801,
  title = {Influence of Nb alloying on Nb recrystallization and the upper critical field of \text{Nb$_{3}$Sn}},
  author = {Paudel, Nawaraj and Tarantini, Chiara and Balachandran, Shreyas and Starch, William L. and Lee, Peter J. and Larbalestier, David C.},
  journal = {Phys. Rev. Mater.},
  volume = {8},
  issue = {8},
  pages = {084801},
  numpages = {14},
  year = {2024},
  month = {Aug},
  publisher = {American Physical Society},
  doi = {10.1103/PhysRevMaterials.8.084801},
  url = {https://link.aps.org/doi/10.1103/PhysRevMaterials.8.084801}
}

@article{Ta-Hf,
  title = {Superconductivity with high upper critical field in \text{Ta-Hf} alloys},
  author = {Meena, P. K. and Jangid, S. and Kushwaha, R. K. and Singh, R. P.},
  journal = {Phys. Rev. Mater.},
  volume = {7},
  issue = {8},
  pages = {084801},
  numpages = {7},
  year = {2023},
  month = {Aug},
  publisher = {American Physical Society},
  doi = {10.1103/PhysRevMaterials.7.084801},
  url = {https://link.aps.org/doi/10.1103/PhysRevMaterials.7.084801}
}

@article{Ta-Zr,
  title = {Superconductivity of \text{Ta-Hf} and \text{Ta-Zr} alloys: Potential alloys for use in superconducting devices},
  author = {Klimczuk, Tomasz and Kr\'olak, Szymon and Cava, Robert J.},
  journal = {Phys. Rev. Mater.},
  volume = {7},
  issue = {6},
  pages = {064802},
  numpages = {10},
  year = {2023},
  month = {Jun},
  publisher = {American Physical Society},
  doi = {10.1103/PhysRevMaterials.7.064802},
  url = {https://link.aps.org/doi/10.1103/PhysRevMaterials.7.064802}
}

@article{NbTipressure,
author = {Guo, Jing and Lin, Gongchang and Cai, Shu and Xi, Chuanying and Zhang, Changjin and Sun, Wanshuo and Wang, Qiuliang and Yang, Ke and Li, Aiguo and Wu, Qi and Zhang, Yuheng and Xiang, Tao and Cava, Robert Joseph and Sun, Liling},
title = {Record-High Superconductivity in Niobium-Titanium Alloy},
journal = {Adv. Mat.},
volume = {31},
number = {11},
pages = {1807240},
doi = {https://doi.org/10.1002/adma.201807240},
url = {https://advanced.onlinelibrary.wiley.com/doi/abs/10.1002/adma.201807240},
year = {2019}
}

@article{lowTcGi,
  title = {Melting of vortex lattice in the magnetic superconductor \text{RbEuFe$_{4}$As$_{4}$}},
  author = {Koshelev, A. E. and Willa, K. and Willa, R. and Smylie, M. P. and Bao, J.-K. and Chung, D. Y. and Kanatzidis, M. G. and Kwok, W.-K. and Welp, U.},
  journal = {Phys. Rev. B},
  volume = {100},
  issue = {9},
  pages = {094518},
  numpages = {8},
  year = {2019},
  month = {Sep},
  publisher = {American Physical Society},
  doi = {10.1103/PhysRevB.100.094518},
  url = {https://link.aps.org/doi/10.1103/PhysRevB.100.094518}
}

@article{TixNb,
  title={Titanium-niobium (\text{Ti-xNb}) alloys with high Nb amounts for applications in biomaterials},
  author={Pereira, Bruno Leandro and Lepienski, Carlos Maur{\'\i}cio and Seba, Viviane and Hobold, Guibert and Soares, Paulo and Chee, Bor Shin and Kuroda, Pedro Akira Bazaglia and Szameitat, Erico Saito and Santos, Leonardo Lu{\'\i}s dos and Grandini, Carlos Roberto and others},
  journal={Materials Research},
  volume={23},
  number={6},
  pages={e20200405},
  year={2020},
  publisher={SciELO Brasil},
  doi={http://dx.doi.org/10.1590/1980-5373-MR-2020-0405}
}

@article{qbits1,
author = {Nathalie P. de Leon  and Kohei M. Itoh  and Dohun Kim  and Karan K. Mehta  and Tracy E. Northup  and Hanhee Paik  and B. S. Palmer  and N. Samarth  and Sorawis Sangtawesin  and D. W. Steuerman },
title = {Materials challenges and opportunities for quantum computing hardware},
journal = {Science},
volume = {372},
number = {6539},
pages = {eabb2823},
year = {2021},
doi = {10.1126/science.abb2823},
URL = {https://www.science.org/doi/abs/10.1126/science.abb2823}
}

@article{qbits2,
  title={New material platform for superconducting transmon qubits with coherence times exceeding 0.3 milliseconds},
  author={Place, Alexander PM and Rodgers, Lila VH and Mundada, Pranav and Smitham, Basil M and Fitzpatrick, Mattias and Leng, Zhaoqi and Premkumar, Anjali and Bryon, Jacob and Vrajitoarea, Andrei and Sussman, Sara and others},
  journal={Nat. comm.},
  volume={12},
  number={1},
  pages={1779},
  year={2021},
  doi={https://doi.org/10.1038/s41467-021-22030-5},
  publisher={Nature Publishing Group UK London}
}

@article{ZrdopedNb,
  title = {Enhanced Surface Superconductivity of Niobium by Zirconium Doping},
  author = {Sitaraman, Nathan S. and Sun, Zeming and Francis, Benjamin L. and Hire, Ajinkya C. and Oseroff, Thomas and Baraissov, Zhaslan and Arias, Tomas A. and Hennig, Richard G. and Liepe, Matthias U. and Muller, David A. and Transtrum, Mark K.},
  collaboration = {Center for Bright Beams},
  journal = {Phys. Rev. Appl.},
  volume = {20},
  issue = {1},
  pages = {014064},
  numpages = {10},
  year = {2023},
  month = {Jul},
  publisher = {American Physical Society},
  doi = {10.1103/PhysRevApplied.20.014064},
  url = {https://link.aps.org/doi/10.1103/PhysRevApplied.20.014064}
}

@article{NbX,
title = {The strong influence of \text{Ti}, \text{Zr}, \text{Hf} solutes and their oxidation on microstructure and performance of \text{Nb3Sn} superconductors},
journal = {J. Alloys Compounds},
volume = {857},
pages = {158270},
year = {2021},
issn = {0925-8388},
doi = {https://doi.org/10.1016/j.jallcom.2020.158270},
url = {https://www.sciencedirect.com/science/article/pii/S0925838820346338},
author = {X. Xu and X. Peng and J. Rochester and M.D. Sumption and J. Lee and G.A. {Calderon Ortiz} and J. Hwang}
}

@article{NbTihon2003composition,
  title={Composition/phase structure and properties of titanium-niobium alloys},
  author={Hon, Yen-Huei and Wang, Jian-Yih and Pan, Yung-Ning},
  journal={Materials transactions},
  volume={44},
  number={11},
  pages={2384--2390},
  year={2003},
  doi={https://doi.org/10.2320/matertrans.44.2384},
  publisher={The Japan Institute of Metals and Materials}
}

@article{coombs2024high,
  title={High-temperature superconductors and their large-scale applications},
  author={Coombs, Tim A and Wang, Qi and Shah, Adil and Hu, Jintao and Hao, Luning and Patel, Ismail and Wei, Haigening and Wu, Yuyang and Coombs, Thomas and Wang, Wei},
  journal={Nat. Rev. Electr. Eng.},
  pages={1--14},
  year={2024},
  doi= {10.1038/s44287-024-00112-y},
  publisher={Nature Publishing Group UK London}
}

@article{Nb3SnPRM,
  title = {Influence of \text{Nb} alloying on \text{Nb} recrystallization and the upper critical field of \text{Nb$_{3}$Sn}},
  author = {Paudel, Nawaraj and Tarantini, Chiara and Balachandran, Shreyas and Starch, William L. and Lee, Peter J. and Larbalestier, David C.},
  journal = {Phys. Rev. Mater.},
  volume = {8},
  issue = {8},
  pages = {084801},
  numpages = {14},
  year = {2024},
  month = {Aug},
  publisher = {American Physical Society},
  doi = {10.1103/PhysRevMaterials.8.084801},
  url = {https://link.aps.org/doi/10.1103/PhysRevMaterials.8.084801}
}

@article{NbZrPRM,
  title = {Structure and coherency of bcc \text{Nb} precipitates in hcp \text{Zr} matrix from atomistic simulations},
  author = {Fan, Zhengxuan and Maras, \'Emile and Cottura, Maeva and Marinica, Mihai-Cosmin and Clouet, Emmanuel},
  journal = {Phys. Rev. Mater.},
  volume = {8},
  issue = {11},
  pages = {113601},
  numpages = {14},
  year = {2024},
  month = {Nov},
  publisher = {American Physical Society},
  doi = {10.1103/PhysRevMaterials.8.113601},
  url = {https://link.aps.org/doi/10.1103/PhysRevMaterials.8.113601}
}

@article{hampshire1974critical,
  title={The critical current-density of \text{Nb-60} at\% \text{Ti} and \text{Nb-25} at\% \text{Zr} superconductors in small magnetic fields},
  author={Hampshire, RG},
  journal={J. Phys. D: Applied Physics},
  volume={7},
  number={13},
  pages={1847},
  year={1974},
  doi={10.1088/0022-3727/7/13/311},
  publisher={IOP Publishing}
}

@article{banno2021high,
  title={High-temperature-tolerable superconducting Nb-alloy and its application to Pb-and Cd-free superconducting joints between \text{NbTi} and \text{Nb$_{3}$Sn} wires},
  author={Banno, Nobuya and Kobayashi, Kensuke and Uchida, Akira and Kitaguchi, Hitoshi},
  journal={Journal of Materials Science},
  volume={56},
  pages={20197--20207},
  year={2021},
  doi={https://doi.org/10.1007/s10853-021-06585-8},
  publisher={Springer}
}

@article{Nb0.75Zr0.25,
  title={Eliashberg analysis of the specific heat of \text{Nb$_{0.75}$Zr$_{0.25}$}},
  author={Junod, A and Jorda, J -L and Muller, J},
  journal={Journal of low temperature physics},
  volume={62},
  pages={301--313},
  year={1986},
  doi={https://doi.org/10.1007/BF00683466},
  publisher={Springer}
}

@article{Nb-Hf,
      title={Precipitates and fluxoid pinning in a superconducting \text{Nb-Hf} alloy},
  author={Koch, CC and Carpenter, RW},
  journal={Philosophical Magazine},
  volume={25},
  number={2},
  pages={303--320},
  year={1972},
  doi={https://doi.org/10.1080/14786437208226807},
  publisher={Taylor \& Francis}
}

@article{fietz1964magnetization,
  title={Magnetization of superconducting \text{Nb}-25\% \text{Zr} wire},
  author={Fietz, WA and Beasley, MR and Silcox, J and Webb, WW},
  journal={Phys. Rev.},
  volume={136},
  number={2A},
  pages={A335},
  year={1964},
  publisher={APS},
  doi = {10.1103/PhysRev.136.A335},
  url = {https://link.aps.org/doi/10.1103/PhysRev.136.A335}
}

@article{NbZrPhysRevLett.6.671,
  title = {Superconductivity at High Magnetic Fields and Current Densities in Some \text{Nb-Zr} Alloys},
  author = {Berlincourt, T. G. and Hake, R. R. and Leslie, D. H.},
  journal = {Phys. Rev. Lett.},
  volume = {6},
  issue = {12},
  pages = {671--674},
  numpages = {0},
  year = {1961},
  month = {Jun},
  publisher = {American Physical Society},
  doi = {10.1103/PhysRevLett.6.671},
  url = {https://link.aps.org/doi/10.1103/PhysRevLett.6.671}
}

@article{NbTiPhysRevB.110.L140502,
  title = {\text{NbTi}: A nontrivial puzzle for the conventional theory of superconductivity},
  author = {Cucciari, Alessio and Naddeo, Dionisia and Di Cataldo, Simone and Boeri, Lilia},
  journal = {Phys. Rev. B},
  volume = {110},
  issue = {14},
  pages = {L140502},
  numpages = {6},
  year = {2024},
  month = {Oct},
  publisher = {American Physical Society},
  doi = {10.1103/PhysRevB.110.L140502},
  url = {https://link.aps.org/doi/10.1103/PhysRevB.110.L140502}
}

@ARTICLE{largascaleapplication,
  author={Scanlan, R.M. and Malozemoff, A.P. and Larbalestier, D.C.},
  journal={Proceedings of the IEEE}, 
  title={Superconducting materials for large scale applications}, 
  year={2004},
  volume={92},
  number={10},
  pages={1639-1654},
  doi={10.1109/JPROC.2004.833673}
}

@article{Kinoshita01081990,
author = {Kyoichi Kinoshita},
title = {Crystal structures and properties of superconducting materials I},
journal = {Phase Transitions},
volume = {23},
number = {2-4},
pages = {73--250},
year = {1990},
publisher = {Taylor \& Francis},
doi = {10.1080/01411599008241820},
URL = {https://doi.org/10.1080/01411599008241820},
}

@article{patel2019niobium,
  title={Niobium-titanium \text{(Nb-Ti)} superconducting joints for persistent-mode operation},
  author={Patel, Dipak and Kim, Su-Hun and Qiu, Wenbin and Maeda, Minoru and Matsumoto, Akiyoshi and Nishijima, Gen and Kumakura, Hiroaki and Choi, Seyong and Kim, Jung Ho},
  journal={Sci. Rep.},
  volume={9},
  number={1},
  pages={14287},
  year={2019},
doi={https://doi.org/10.1038/s41598-019-50549-7},
  publisher={Nature Publishing Group UK London}
}

@article{mirmefstein1997mixed,
  title={Mixed-State Specific Heat of the Type-II Superconductor \text{Nb$_{0.77}$Zr$_{0.23}$} in Magnetic Fields up to \text{B$_{c2}$}},
  author={Mirmefstein, A and Junod, A and Walker, E and Revaz, B and Genoud, JY and Triscone, G},
  journal={Journal of superconductivity},
  volume={10},
  pages={527--535},
  year={1997},
  doi={https://doi.org/10.1007/BF02767690},
  publisher={Springer}
}

@article{baker1969correlation,
  title={Correlation of superconducting and metallurgical properties of a \text{Ti-20} at.\% \text{Nb} alloy},
  author={Baker, C and Sutton, J},
  journal={Philosophical Magazine},
  volume={19},
  number={162},
  pages={1223--1255},
  year={1969},
  doi= {https://doi.org/10.1080/14786436908228647},
  publisher={Taylor \& Francis}
}

@article{priinits2024peculiarities,
  title={Peculiarities of niobium-based superconducting alloys in the light of crystal chemistry: A brief survey},
  author={Priinits, Taimo and Vargunin, Artjom and Liivand, Aleksandr},
  journal={arXiv:2406.00817},
  year={2024},
  url={https://arxiv.org/pdf/2406.00817v1}
}

@misc{jablonski1993niobium,
  title={Niobium-Titanium superconductors produced by powder metallurgy having artificial flux pinning centers},
  author={Jablonski, Paul D and Larbalestier, David C},
  year={1993},
  month={jul},
  publisher={Google Patents},
  note={\text{US} Patent 5,226,947}
}

@article{rodrigues2000development,
  title={Development of \text{(Nb,Ta)$_{3}$Sn} multifilamentary superconductor wire for high current applications},
  author={Rodrigues Jr, Durval and Machado, Jo{\~a}o Paulo Barros},
  journal={Mat. Res.},
  volume={3},
  pages={99--103},
  year={2000},
  doi={https://doi.org/10.1590/S1516-14392000000400002},
  publisher={SciELO Brasil}
}

@article{STEWART201528,
title = {Superconductivity in the \text{A15} structure},
author = {G.R. Stewart},
journal = {Physica C},
volume = {514},
pages = {28-35},
year = {2015},
note = {Superconducting Materials: Conventional, Unconventional and Undetermined},
issn = {0921-4534},
doi = {https://doi.org/10.1016/j.physc.2015.02.013},
url = {https://www.sciencedirect.com/science/article/pii/S0921453415000404}
}

@incollection{junod2002specificNb0.77Zr0.23,
  title={Specific Heat Experiments in High Magnetic Fields: D-Wave Symmetry, Fluctuations, Vortex Melting},
  author={Junod, Alain and Roulin, Marlyse and Revaz, Bernard and Erb, Andreas and Walker, Eric},
  booktitle={The Gap Symmetry and Fluctuations in High-Tc Superconductors},
  pages={403--421},
  year={2002},
  publisher={Springer},
  doi= {https://doi.org/10.1007/0-306-47081-0_22}
}

@InProceedings{Nb-Hfconf,
author={Bychkova, M. I. and Baron, V. V. and Savitskii, E. M.},
editor={Savitskii, E. M. and Baron, V. V.},
title={Superconducting Properties of \text{Nb-Hf} Alloys and the Effect of Mechanical and Heat Treatment on Their Structure and Properties},
booktitle={Physics and Metallurgy of Superconductors},
year={1970},
publisher={Springer US},
pages={74--79},
doi={10.1007/978-1-4684-8220-1_13}
}

@article{Nb-Hfkoch1979peak,
  title={The peak effect, summation problem, and magnetic history in a superconducting \text{Nb-38} at.\% |text{Hf} alloy},
  author={Koch, CC and DasGupta, A and Kroeger, DM and Scarbrough, JO},
  journal={Philosophical Magazine B},
  volume={40},
  number={5},
  pages={361--387},
  year={1979},
  doi={https://doi.org/10.1080/13642817908246379},
  publisher={Taylor \& Francis}
}

@article{seeber1998commercially,
  title={Commercially available superconducting wires},
  author={Seeber, B},
  journal={Handbook of Applied Superconductivity},
  pages={397--488},
  year={1998},
  publisher={Inst. of Physics}
}

@misc{seeber1998power,
  title={Power Applications of Superconductivity: Handbook of Applied Superconductivity. Bristol, UK: Inst},
  author={Seeber, B},
  year={1998},
  publisher={Physics}
}

@inproceedings{kalsi2003installation,
  title={Installation and operation of superconducting rotating machines},
  author={Kalsi, Swam S},
  booktitle={Transmission and Distribution Conference and Exposition},
  volume={3},
  pages={1098--1101},
  year={2003},
  organization={IEEE},
  doi={10.1109/TDC.2003.1335103}
}

@article{Nb3Al_recent,
  title={New-route of flux pinning enhancement in \text{Nb$_3$Al} superconductor: density and orientation regulation of stacking faults by nano-oxide particles},
  author={Yang, Changkun and Yu, Zhou and Li, Wenlong and Zhang, Huaihao and Ren, Jiahao and Wang, Wentao and Zhang, Yong and Zhao, Yong},
  journal={Supercond. Sci. Technol.},
  year={2025},
  url={10.1088/1361-6668/adaf5d}
}

@article{PhysRevB.85.174505,
  title = {Superconductivity in the Heusler family of intermetallics},
  author = {Klimczuk, T. and Wang, C. H. and Gofryk, K. and Ronning, F. and Winterlik, J. and Fecher, G. H. and Griveau, J.-C. and Colineau, E. and Felser, C. and Thompson, J. D. and Safarik, D. J. and Cava, R. J.},
  journal = {Phys. Rev. B},
  volume = {85},
  issue = {17},
  pages = {174505},
  numpages = {8},
  year = {2012},
  month = {May},
  publisher = {American Physical Society},
  doi = {10.1103/PhysRevB.85.174505},
  url = {https://link.aps.org/doi/10.1103/PhysRevB.85.174505}
}

@article{gurevich2006enhancement,
  title={Enhancement of rf breakdown field of superconductors by multilayer coating},
  author={Gurevich, Alexander},
  journal={Appl. Phys. Let.},
  volume={88},
  number={1},
  year={2006},
  publisher={AIP Publishing},
  url = {https://doi.org/10.1063/1.2162264}
}

@article{valente2016superconducting,
  title={Superconducting \text{RF} materials other than bulk niobium: a review},
  author={Valente-Feliciano, Anne-Marie},
  journal={Supercond. Sci. Technol.},
  volume={29},
  number={11},
  pages={113002},
  year={2016},
  publisher={IOP Publishing},
  URL = {10.1088/0953-2048/29/11/113002}
}

@article{bean1964magnetization,
  title={Magnetization of high-field superconductors},
  author={Bean, Charles P},
  journal={Reviews of modern physics},
  volume={36},
  number={1},
  pages={31},
  year={1964},
  publisher={APS}
}

@article{PhysRevB.75.134515,
  title = {Flux pinning force in bulk $\mathrm{Mg}{\mathrm{B}}_{2}$ with variable grain size},
  author = {Mart\'{\i}nez, E. and Mikheenko, P. and Mart\'{\i}nez-L\'opez, M. and Mill\'an, A. and Bevan, A. and Abell, J. S.},
  journal = {Phys. Rev. B},
  volume = {75},
  issue = {13},
  pages = {134515},
  numpages = {8},
  year = {2007},
  month = {Apr},
  publisher = {American Physical Society},
  doi = {10.1103/PhysRevB.75.134515},
  url = {https://link.aps.org/doi/10.1103/PhysRevB.75.134515}
}

@phdthesis{He2020,
  title={Radiation damage and its impact on corrosion in Zirconium-Niobium alloys},
  author={He, Guanze},
  year={2020},
  school={University of Oxford}
}

@article{Goel2021,
  title={Study on the dissolution of $\beta$-precipitates in the Zr--Nb alloy under the influence of Ne ion irradiation},
  author={Goel, Lokesh and Mir, Anamul H and Naveen Kumar, N and Satyam, Parlapalli V and Hinks, Jonathan A and Donelly, Stephen E and Tewari, Raghvendra},
  journal={Microscopy},
  volume={70},
  number={5},
  pages={461--468},
  year={2021},
  publisher={Oxford University Press UK}
}

@article{Azevedo2011,
  title={Selection of fuel cladding material for nuclear fission reactors},
  author={Azevedo, CR de F},
  journal={Engineering Failure Analysis},
  volume={18},
  number={8},
  pages={1943--1962},
  year={2011},
  publisher={Elsevier}
}

@article{Shapira1965,
  title = {Upper Critical Fields of Nb-Ti Alloys: Evidence for the Influence of Pauli Paramagnetism},
  author = {Shapira, Y. and Neuringer, L. J.},
  journal = {Phys. Rev.},
  volume = {140},
  issue = {5A},
  pages = {A1638--A1644},
  numpages = {0},
  year = {1965},
  month = {Nov},
  publisher = {American Physical Society},
  doi = {10.1103/PhysRev.140.A1638},
  url = {https://link.aps.org/doi/10.1103/PhysRev.140.A1638}
}

@article{fluxjumpsinNb,
  title = {Origin of magnetic flux-jumps in Nb films subject to mechanical vibrations and corresponding magnetic perturbations},
  author = {Golovchanskiy, Igor A. and Pan, Alexey V. and Johansen, Tom H. and George, Jonathan and Rudnev, Igor A. and Rosenfeld, Anatoly and Fedoseev, Sergey A.},
  journal = {Phys. Rev. B},
  volume = {97},
  issue = {1},
  pages = {014524},
  numpages = {7},
  year = {2018},
  month = {Jan},
  publisher = {American Physical Society},
  doi = {10.1103/PhysRevB.97.014524},
  url = {https://link.aps.org/doi/10.1103/PhysRevB.97.014524}
}

@article{Zhao2021NbTi,
  author    = {In Yong Moon and Se-Jong Kim and Ho Won Lee and Jaimyun Jung and Young-Seok Oh and Seong-Hoon Kang},
  title     = {Investigation of the Correlation between Initial Microstructure and Critical Current Density of Nb-46.5 wt\%Ti Superconducting Material},
  journal   = {Metals},
  year      = {2021},
  volume    = {11},
  number    = {5},
  pages     = {777},
  doi       = {10.3390/met11050777},
  url       = {https://www.mdpi.com/2075-4701/11/5/777}
}

@article{Xu2021Nb3Sn,
  author    = {Xu, Xingchen and Peng, X. and Rochester, J. and Sumption, M. and Lee, J. and Calderon Ortiz, G. and Hwang, J.},
  title     = {The strong influence of Ti, Zr, Hf solutes and their oxidation on microstructure and performance of Nb3Sn superconductors},
  journal   = {Journal of Alloys and Compounds},
  year      = {2021},
  volume    = {857},
  pages     = {158270},
  doi       = {10.1016/j.jallcom.2020.158270}
}

@misc{Eisterer2018Irradiation,
  author    = {Eisterer, M.},
  title     = {Neutron Irradiation: Introduced Defects in Various Superconductors},
  howpublished = {IEEE CSC Presentation},
  year      = {2018},
  note      = {Conference presentation}
}

@article{Thompson1979Nb3Ge,
  author    = {Thompson, J. D. and Maley, M. P. and Newkirk, L. R. and Bartlett, R. J.},
  title     = {High-field properties and scaling in CVD-prepared Nb3Ge},
  journal   = {Journal of Applied Physics},
  year      = {1979},
  volume    = {50},
  number    = {2},
  pages     = {977--982},
  doi       = {10.1063/1.326021}
}

@article{Webb1973Nb3Ge,
  author    = {Webb, G. W. and Tainsh, R. J. and Dynes, R. C.},
  title     = {Superconducting properties of sputtered Nb3Ge films},
  journal   = {Physical Review Letters},
  year      = {1973},
  volume    = {30},
  pages     = {401--404},
  doi       = {10.1103/PhysRevLett.30.401}
}

@article{Flukiger1981Nb3Ga,
  author    = {Fl{\"u}kiger, R. and Devantay, H. and M{\"u}ller, J.},
  title     = {Superconducting properties of Nb3Ga prepared by different techniques},
  journal   = {IEEE Transactions on Magnetics},
  year      = {1981},
  volume    = {17},
  number    = {1},
  pages     = {331--334},
  doi       = {10.1109/TMAG.1981.1060792}
}

@article{Joshi2018NbN,
  author    = {Joshi, L. M. and Verma, A. and Gupta, A. and Rout, P. K. and Husale, S. and Budhani, R. C.},
  title     = {Superconducting properties of NbN film, bridge and meanders},
  journal   = {AIP Advances},
  year      = {2018},
  volume    = {8},
  number    = {5},
  pages     = {055305},
  doi       = {10.1063/1.5026219}
}

@article{Jing2023NbN,
  author    = {Jing, T. Y. and Han, Z. Y. and He, Z. H. and Shao, M. X. and Li, P. and Li, Z. Q.},
  title     = {Quantum phase transition in NbN superconducting thin films},
  journal   = {Physical Review B},
  year      = {2023},
  volume    = {107},
  number    = {18},
  pages     = {184515},
  doi       = {10.1103/PhysRevB.107.184515}
}

@inproceedings{AIP2023NbN,
  author    = {Anonymous},
  title     = {Enhancement of critical current density in superconducting NbN thin film using vortex-matter manipulation},
  booktitle = {AIP Conference Proceedings},
  year      = {2023},
  volume    = {2778},
  pages     = {030008},
  doi       = {10.1063/5.0164825}
}
\end{document}